\documentclass[10pt]{article}
\usepackage[a4paper,left=3cm,right=3cm]{geometry}

\usepackage[utf8]{inputenc}
\usepackage[T1]{fontenc}
\usepackage{tikz}

\usepackage{amsmath, amssymb,amsthm}
\usepackage{authblk}
\usepackage{booktabs}
\usepackage{url}
\usepackage{bm}
\usepackage{graphicx}
\usepackage{enumerate}
\usepackage{placeins}
\usepackage{caption}
\usepackage[labelformat=simple]{subcaption}
\usepackage{tablefootnote}
\usepackage{mathtools}

\theoremstyle{definition}

\theoremstyle{remark}

\numberwithin{equation}{section}

\newcommand{\RR}{\mathbb{R}}

\newcommand{\cL}{\mathcal{L}}

\newcommand{\set}[1]{\left\{#1\right\}}

\newcommand{\ind}[1]{\bm{1}_{#1}}

\newcommand{\Sym}{\mathsf{Sym}}

\begin{document}
\pagestyle{empty}

\title{Discovering parametrizations of implied volatility with symbolic regression}
\author{Martin Keller-Ressel}
\author{Hannes Nikulski}
\affil{TU Dresden, Institute for Mathematical Stochastics, Dresden, 01062, Germany}

\maketitle

\begin{abstract}
We investigate the data-driven discovery of parametric representations for implied volatility slices. Using symbolic regression, we search for simple analytic formulas that approximate the total implied variance as a function of log-moneyness and maturity. Our approach generates candidate parametrizations directly from market data without imposing a predefined functional form. We compare the resulting formulas with the widely used SVI parametrization in terms of accuracy and simplicity. Numerical experiments indicate that symbolic regression can identify compact parametrizations with competitive fitting performance.
\end{abstract}

\section{Introduction}
Building and maintaining an implied volatility surface from option quotes is one of the key tasks of market makers in option markets. This task includes inter- and extrapolating implied volatilities (IVs), detecting invalid quotes and being able to generate realistic scenarios of IV surfaces. To do so, parametric methods, such as the SABR approximation of \cite{hagan2002managing} or the SVI method of \cite{gatheral2014arbitrage} are frequently used, although non- or semi-parametric methods \cite{fengler2005semiparametric} as well as ML-based methods, such as AEs and VAEs \cite{vuletic2024volgan} also exist. Our point of reference is the SVI (Stochastic Volatility Inspired) method, which parametrizes \emph{total implied variance} $w(k)$ of an option with log-moneness $k$ as
\begin{equation}\label{eq:SVI}
w(k) = a + b \left\{\rho (k - m) + \sqrt{(k-m)^2 +s^2}\right\},
\end{equation}
where $a \in \RR, b \ge 0, |\rho| < 1, m \in \RR$ and $s > 0$. The key advantages of the SVI are its simplicity, interpretability and empirical accuracy, i.e., its demonstrated ability to typically provide good fits to observed implied volatilities. Here, we are interested in the following question: 
\begin{quote}
Are there other simple parametrizations of implied volatility with similar or better accuracy than SVI?
\end{quote}
We aim to discover these alternative parametrizations in a purely data-driven fashion, using \emph{symbolic regression}. Symbolic regression is a method to find a functional relationship
\[f(x_i) \approx y_i \]
between input data $x_i$ and responses $y_i$. Contrary to other types of regression, where $f$ is assumed to have a predefined form (linear, logistic, spline, etc.), determining the functional form of $f$ is part of the symbolic regression problem. In doing so, $f$ can take the form of an arbitrary algebraic expression, subject to generous constraints. As an example, using empirical data on planetary orbits (consisting of length of semi-major axis and orbital period), symbolic regression has been shown to be able to rediscover Kepler's Third Law of planetary motion, i.e. the functional relationship $(\text{period})^2 \propto (\text{radius})^3$, see \cite{cranmer2023interpretable}.\\

The symbolic regression problem for implied volatilities is a non-standard problem in several aspects: First, implied volatilities do not constitute a standard (`independent and identically distributed') statistical sample. Instead, option pricing data is highly structured by its dependency on strike price and time-to-maturity of the option. Accordingly, implied volatilities are usually grouped into \emph{slices} (implied volatilities of the same maturity, indexed by strike price) and this structure has to be considered in the setup of the regression problem.  Second, implied volatilities are subject to no-arbitrage constraints, whose highly non-linear nature has been subject of several publications \cite{roper2009implied, doust2012no, martini2022no}. Reasonably, such constraints should be included in the symbolic regression problem. \\

We describe the background and our methods in Section~\ref{sec:methods} and report our results on two different data sets in Sections~\ref{sec:D1} and \ref{sec:D2}. We conclude with a short discussion of our findings in Section~\ref{sec:conclusion}.

\section{Background and Methods}\label{sec:methods}

\subsection{Implied Volatility Curves and No-Arbitrage Constraints}\label{sec:arbitrage1}
Implied volatility data is typically given as a structured data set
\begin{equation}\label{eq:structure}
(x_{ij}, y_{ij}), \qquad i=1, \dotsc, N_\text{slice}, \quad j=1, \dotsc, M(i),
\end{equation}
where each $i$ represents a `slice' of option pricing data, i.e. data for options of a specific maturity quoted on a specific trading day, and $j$ cycles through all $M(i)$ options of different moneyness within a specific slice. The input $x_{ij} = \log(K/S)$ is log-moneyness, derived from the option's strike price $K$ and spot price $S$. The target $y_{ij} = \sigma^2 \tau$ is total implied variance, derived from the option's implied volatility $\sigma$ and its time-to-maturity $\tau$. Our objective is to find a family $f(.,p)$ of functions, parametrized by a parameter vector $p \in \RR^d$, which for each slice $i$ gives a good approximation 
\begin{equation}
f(x_{ij}, p_i) \approx y_{ij},
\end{equation}
after fitting the parameter $p_i$. For the SVI method shown in \eqref{eq:SVI}, we have $d = 5$ and $p = (a,b,\rho,m,s)$. It is well-known that option prices, and, by extension, also implied volatilities, must satisfy certain constraints, in order to avoid arbitrage in the options' market, see \cite{roper2009implied}. For a single slice of IVs, these constraints are\footnote{The three listed constraints are necessary for absence of arbitrage and are very close to being sufficient; see \cite{roper2009implied} for details}: 
\begin{enumerate}[(a)]
\item Positivity of total implied volatility: $y(x) > 0\;\forall x\in \RR$;
\item Durrleman's condition: $y$ is twice differentiable, and $g(x) \geq 0\;\forall x\in \RR$, where 
    \begin{equation}
        \label{eq:g}
        g(x) := \left(1 - \frac{x\partial_x y(x)}{2y(x)}\right)^2 - \frac{\partial_x y\left(x\right)^2}{4}\left( \frac{1}{y(x)} + \frac{1}{4} \right) + \frac{\partial_{xx}y(x)}{2}.
    \end{equation}
\item Lee's tail bounds \cite{lee2004moment}:
\[\limsup_{x \to -\infty} \frac{y(x)}{|x|} \le 2, \quad \text{and} \quad \limsup_{x \to \infty} \frac{y(x)}{x} \le 2\]
\end{enumerate}
Durrleman's condition is closely related to the absence of butterfly arbitrage (arbitrage between options with different strikes, but the same maturity), cf. \cite{gatheral2014arbitrage}. The limiting slopes that appear in Lee's tail bounds can be linked to the tail behaviour of the risk-neutral spot price density, see \cite{lee2004moment}, and are hence related to asymptotic arbitrage in the wings of the IV smile.


\subsection{Symbolic Regression with \texttt{PySR}}
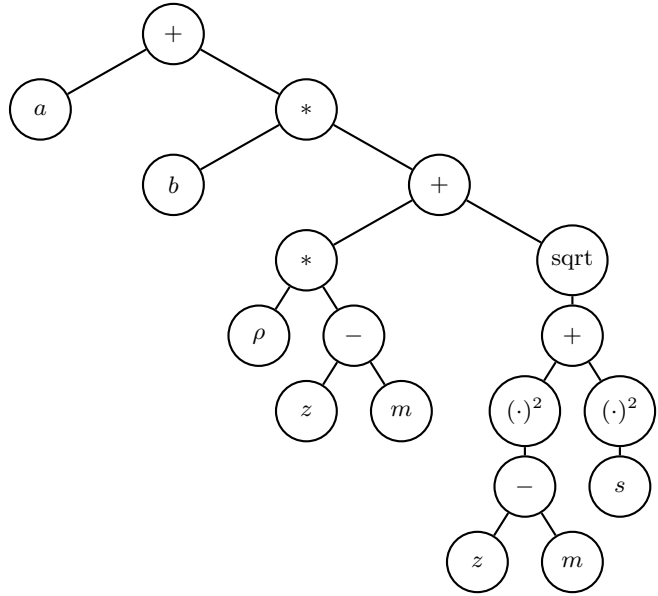
\begin{figure}[htbp]
    \centering

\def\borderColor{color2}
\tikzset{
    every node/.style={
            draw,
            thick,
            circle,
            text=color1,
            color=black,
            minimum width=0.8cm,
            font=\small,
            align=center
        },
    level distance=1.0cm,
    level 1/.style={sibling distance=3.5cm},
    level 2/.style={sibling distance=3.5cm},
    level 3/.style={sibling distance=3.5cm},
    level 4/.style={sibling distance=1.25cm},
    level 5/.style={sibling distance=1.25cm},
    level 6/.style={sibling distance=1.25cm},
    level 7/.style={sibling distance=1.25cm},
    blank/.style={draw=none},
    edge from parent/.style={
            draw,
            thick,
            color=black,
            edge from parent path={(\tikzparentnode) -- (\tikzchildnode)}
        }
}
\tikzset{boundaryStyle/.style={\borderColor, line width=0.5mm}}
\begin{tikzpicture}
    \node (n1) at (0,0) {$+$}
    child {node {$a$}}
    child {node {$*$}
            child {node {$b$}}
            child {node {$+$}
                    child {node {$*$}
                            child {node {$\rho$}}
                            child {node {$-$}
                                    child {node {$z$}}
                                    child {node {$m$}}
                                }
                        }
                    child {node {$\operatorname{sqrt}$}
                            child {node {$+$}
                                    child {node {$(\cdot)^2$}
                                            child {node {$-$}
                                                    child {node {$z$}}
                                                    child {node {$m$}}
                                                }
                                        }
                                    child {node {$(\cdot)^2$}
                                            child {node {$s$}}
                                        }
                                }
                        }
                }
        };
\end{tikzpicture}
    \caption{Expression tree of the SVI parametrization \eqref{eq:SVI}.}
    \label{fig:svi-tree}
\end{figure}

Given our input space $\RR^{d+1}$ and output space $\RR$, we write $\Sym$ for the set of all valid symbolic expressions representing functions $f = f(x,p)$ from $\RR \times \RR^d$ to $\RR$. A symbolic expression is valid if it is composed of constants and allowed basis functions, and if it respects all additional constraints (on nesting of operations, on maximal complexity, etc.). Commonly used basis functions in symbolic regression are arithmetic operators such as $\left\{ +, -, *, \div\right\}$ or exponential, trigonometric and hyperbolic functions. For our experiments, we use the \texttt{PySR} framework \cite{cranmer2023interpretable}, an open-source Python package for scientific symbolic regression developed by Miles Cranmer and other contributors. In \texttt{PySR}, the set of admissible basis functions is completely customizable and even user-defined functions can be included. Symbolic expressions composed from these basis functions can be represented by an expression tree, whose leaves are the input $x$ and the function's parameters $p$, and whose interior nodes represent the application of basis functions. We show the expression tree for the SVI parametrization in Figure \ref{fig:svi-tree}. The \textit{complexity} of an expression tree is defined as the sum of the complexities of its nodes. \texttt{PySR} sets the complexity of each node to $1$ by default, but custom complexities for different basis functions can also be used. With default settings, the complexity of the SVI parametrization as given in \eqref{eq:SVI}, is $18$. Reparametrizing it as
\begin{equation}
    \label{eq:svi-reparam}
    f^\mathrm{SVI}(x, p) = p_1 + p_2 x + p_3 \sqrt{(x - p_4)^2 + p_5}
\end{equation}
its complexity can be reduced to $15$.

Given a dataset $D = (x_i, y_i)_{i=1}^N$ and a loss function $\cL: \RR^2 \to [0,\infty)$, symbolic regression aims to solve the empirical risk minimization problem 
\begin{equation}\label{eq:sym_min}
\min_{f \in \Sym, p \in \RR^d} \frac{1}{N} \sum_{i=1}^N \cL(f(x_i,p),y_i)
\end{equation}
under constraints on the complexity of the expression $f$. In \texttt{PySR} this minimization is performed in an outer loop and an inner loop. In the outer loop, an evolutionary algorithm modifies the expression tree in order to explore the search space and find expressions of small loss. In the inner loop the parameters $p$ (and possible other constants appearing in the expression tree) are optimized by a conventional numerical optimization method, such as BFGS. The return value of \texttt{PySR}'s symbolic regression routine is then a \emph{leaderboard} $(c_k, f_k)_{k=1}^{C_\text{max}}$ consisting of a range of complexities $c_k$ and the best function $f_k$ of complexity $c_k$ discovered during the search.

In addition, \texttt{PySR} also provides two core functionalities which drastically improve its flexibility.\footnote{See also \texttt{PySR documentation at \url{https://astroautomata.com/PySR/v1.5.9/}}} The first are \emph{template expressions}. They allow the user to define exactly how multiple functions are combined into a target function. Take, for example, the expression $f(x) = \log(1 + g(x))/(2 + h(x))$. The algorithm would search for the subexpressions $g(x)$, and $h(x)$ such that the combined expression $f(x)$minimizes the loss on the provided data.

The second functionality is the ability to search for \emph{conditional parametric expressions}. For that, the input data is divided into different categories. Within each category, the unknown target function is presumed to have the same parameters. Between the categories, the parameters can vary. The algorithm then searches for a single parametric expression $f(x,p)$, but optimises the parameters $p$ of the candidate expressions for each category separately.

\subsection{Symbolic Regression Applied to Implied Volatility Curves}
Recall that in the context of implied volatility, the data set \eqref{eq:structure} is structured into slices. Our objective is to use symbolic regression to find a parametrized family $f(.,p)$ of functions, which for each slice $i$ gives a good approximation 
\[ f(x_{ij}, p_i) \approx y_{ij},\]
after `tuning' the parameter $p_i$. This puts us in the setting of conditional parametric symbolic regression as described above. The `categories' of our data are the IV slices and while the symbolic expression $f(x,p)$ is global, the parameter $p_i$ is specific to each slice. Separating the optimization problem into inner and outer part, we can write the global loss $L_\text{glob}$ as follows:
\begin{equation}\label{eq:error}
L_\text{glob} = \underbrace{\min_{f \in \Sym} \frac{1}{N_\text{slice}}\sum_{i=1}^{N_\text{slice}} \underbrace{\min_{p_i \in \RR^d} \left(\frac{1}{M(i)} \sum_{j=1}^{M(i)} \cL(f(x_{ij}, p_i), y_{ij})\right)}_{\text{numerical fit to slice $i$}}}_{\text{symbolic fit to whole data}}.
\end{equation}
For the local loss $\cL$ we make the natural choice $\cL(x,y) = (x-y)^2$, i.e., the squared deviation between predictions and targets.

\subsection{Including No-Arbitrage Constraints in Symbolic Regression}\label{sec:arbitrage2}
A further step is to include, at least partially, the arbitrage constraints listed in Section~\ref{sec:arbitrage1}. The first constraint (positivity) is typically superfluous and automatically enforced by fitting to positive implied volatilities. The second one (Durrleman's condition) is difficult to implement, as it involves derivatives of $y$ and imposes a large additional computational workload. Therefore, we implement only the third condition (Lee's tail bounds), by testing for its violation (replacing the limits at infinity by larger, but finite values) and including a penalty term 
\[P_i^\text{Lee}(\lambda) = \begin{cases} \lambda & \qquad \text{Lee's tail bounds violated for slice $i$} \\ 1 & \qquad \text{Lee's tail bounds satisfied for slice $i$,} \end{cases}\]
where penalization strength is controlled by $\lambda \gg  1$. The penalized global loss becomes
\begin{equation}\label{eq:error_pen}
L'_\text{glob} =\min_{f \in \Sym} \frac{1}{N_\text{slice}}\sum_{i=1}^{N_\text{slice}}\left\{ \min_{p_i \in \RR^d} \left(\frac{1}{M(i)} \sum_{j=1}^{M(i)} \cL(f(x_{ij}, p_i), y_{ij})\right) \cdot P_i^\text{Lee}(\lambda) \right\}.
\end{equation}
To monitor Durrleman's condition, which is not directly penalized, we introduce the indicator
\[\alpha_K := \frac{\int_{-K}^K |g(x)| \ind{g(x) < 0} dx}{\int_{-K}^K |g|(x) dx} \ge 0,\]
where $g$ is given by \eqref{eq:g}, and where $K$ is set to a large value, such that $[-K,K]$ contains all practically relevant values of log-moneyness $x$. The indicator $\alpha$ measures the area of the `pseudo-density' $g$ below the $x$-axis and between the bounds $[-K,K]$, relative to the total area both below and above the $x$-axis. If no violation of Durrleman's condition takes place, $\alpha_K$ should be equal to zero for all $K > 0$.

\subsection{Reducing Computational Load}\label{sec:clustering}

Although \texttt{PySR} offers the functionality of solving the conditional parametric symbolic regression problem \eqref{eq:error_pen}, its runtime is very sensitive to the number $N_\text{slice}$ of categories in the data and strongly deteriorates with larger numbers of categories. Our main data set contains 61428 separate IV slices, a number of categories that is vastly unfeasible for the conditional regression facility of \texttt{PySR}. A potential solution is subsampling, i.e., to minimize \eqref{eq:error_pen} only over a random subsample (a `batch') of slices, possibly choosing a different subsample in each step of the optimization algorithm. Indeed, \texttt{PySR} also offers such a batching option, but even with batching activated, the global problem remains infeasible. Instead, we opted for a different approximation heuristic: Using $k$-means clustering, we clustered the IV slices into $N_\text{clust} \in \set{10,30}$ clusters and then used the cluster centers $(\bar x_{cj}, \bar y_{cj})_{c=1}^{N_\text{clust}}$ as prototypical IV slices, resulting in the minimization problem
\begin{equation}\label{eq:error_clus}
L'_\text{clust} = \min_{f \in \Sym} \frac{1}{N_\text{slice}}\sum_{c=1}^{N_\text{clust}}\left\{ \min_{p_c \in \RR^d} \left(\frac{1}{M(c)} \sum_{j=1}^{M(c)} \cL(f(\bar x_{cj}, p_c), \bar y_{cj})\right) \cdot P_c^\text{Lee}(\lambda) \right\}.
\end{equation}
After optimization, we evaluate and compare the proposed expressions using the (unpenalized) global error $L_\text{glob}$;  the clustered error $L'_\text{clust}$ only serves as a surrogate for optimization.

\subsection{Overfitting}
A well-known problem in machine learning is overfitting. Also in symbolic regression, overfitting could in principle take place, i.e. we could find expressions that fit well on the given data set but do not generalize well to additional data. For this reason, we performed several runs of symbolic regression with a train-test-split of $60-40$. We observed that there was virtually no loss difference between the test and the training set and concluded that the complexity penalty imposed by PySR already serves as a suitable regularizer, preventing overfitting. Therefore, further runs were performed on the full data set without train-test-split.

\subsection{Data Sets}
For our experiments we use two data sets, D1 and D2. The first data set D1 consists of SPX implied volatility data covering a period from the 3rd of January 2005 to the 6th of December, 2023 obtained from \url{IVolatility.com}. Between these two dates, we have 4728 usable days of option data, consisting of 61428 separate IV slices. The second data set D2 is a curated data set consisting of 34 implied volatility slices extracted\footnote{Using \texttt{WebPlotDigitizer} from \url{automeris.io}.} from examples of unusual or otherwise remarkable volatility smiles published on the website of \texttt{VolaDynamics} \cite{voladynamics}.

\section{Results on Data Set D1}\label{sec:D1}
We performed sixteen runs of parametric symbolic regression with clustering, as described in Section~\ref{sec:clustering}, using slightly different settings for each run. We always included the following basis functions:  
$$\{+, *, \div, \operatorname{pow}, \operatorname{sqrt}, \operatorname{square}, \operatorname{exp}, \operatorname{log}\}$$
and for some runs also included the hyperbolic tangent ($\operatorname{tanh}$). As templates for $f$ we used either no template, the linear template $f(x,p) = p_1 + p_2 x + g(x,p)$  or the `generalized SVI template' $f(x,p) = p_1 g(x h(p_1), p)$ adapted from \cite[Eq.~(2.3)]{guo2016generalized}. We imposed some restrictions on nesting of functions to avoid spurious combinations like $\exp(\log(x))$ and added a wing-arbitrage penalty, as described in Section~\ref{sec:arbitrage2}. Even for runs with identical settings, we observed different results due to the different initialization of the optimization algorithm of \texttt{PySR}. For each of the runs we stored the leaderboard with complexity capped at $20$, the global loss distribution, and the optimal parameter vector $p_i$ for each slice $i$. 

\begin{figure}[htbp]
    \centering

    \includegraphics[height=8cm]{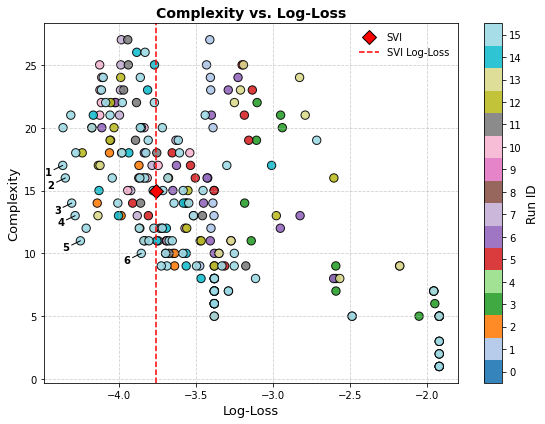}
    \caption{Complexity vs. log-loss of symbolic regression results and comparison to SVI. The six expressions $f_1, \dots f_6$ on the efficient frontier with smaller loss than SVI are marked.} 
    \label{fig:efficient_frontier}
\end{figure}

\begin{table}
\centering
\small
\begin{tabular}{@{}l r r r r @{}}
\toprule
Expression & Complexity & Active Par.  & Log-Loss &  Arb. Ind. $\alpha_K$ \\
\midrule
$f_1(x) = p_3 \sqrt{\tanh(p_1 + x) + p_2}$	& 10	& 3 & -3.855	& 0.0594	\\
$f_2(x) = \log\left(p_3 + \left(\tanh(p_1x) + p_2\right)^2\right)$ & 11& 3 & 	-4.251 & 	0.0986\\
$f_3(x) = \frac{1}{2}\log\left(p_3 + \left(\tanh(p_1x) + p_2\right)^2\right) + p_4 $ & 13	& 4 & -4.286 &	0.0793\\
$f_4(x) = p_4 \left(\tanh(p_1 x) + p_2\right) (p_3 + x) + p_5$ & 14 & 5 & -4.308	& 0.0460\\
$f_5(x) = p_4 \left(\tanh(p_1 x + p_2)\right) (p_3 + x) + p_5$ & 16 & 5 & -4.350	& 0.0821\\
$f_6(x) = p_4 \left(\tanh(p_1 x + p_2^2) + p_2\right) (p_3 + x) + p_5$  & 17& 5 & -4.366	& 0.0770\\
\midrule
$f_\text{SVI}(x) =  p_1 + p_2 x + p_3 \sqrt{(x - p_4)^2 + p_5}$ & $15$ & $5$ & $-3.763$ & $0.0447$\\
\bottomrule
\end{tabular}
\caption{Expressions discovered by symbolic regression, which have lower loss than SVI and are located on the efficient frontier of the complexity vs. log-loss tradeoff in Figure~\ref{fig:efficient_frontier}.}
\label{tab:expr}
\end{table}

In Figure~\ref{fig:efficient_frontier} we show a scatter plot of complexity vs. logarithmic loss over all leaderboard expressions from all sixteen runs of the symbolic regression algorithm. We observe that symbolic regression is able to discover many expressions that are superior to SVI in terms of the complexity vs. accuracy tradeoff. The six expressions on the efficient frontier of the complexity vs. log-loss tradeoff with smaller loss than SVI are marked, and we list the corresponding expressions $f_1, \dotsc, f_6$ in Table~\ref{tab:expr}.\footnote{For the table, we have edited the expressions returned by \texttt{PySR} slightly for interpretability. This includes renumbering of parameters, simple complexity-preserving reparametrizations (like replacing $\frac{x}{p}$ by $xp'$) or performing small mathematical simplifications. The complexities listed are the complexities of the original expression returned by \texttt{PySR}.} Expressions $f_1, f_4, f_5$ and $f_6$ come from the same run of the regression algorithm and their structural similarity is explained by the mutation/crossover method of \texttt{PySR} that explores the search space of valid expressions. While all expressions have a slightly larger arbitrage indicator $\alpha_K$ than SVI, the difference is not dramatic. We highlight expression $f_4$ which has lower complexity and lower loss than SVI, and an arbitrage indicator that is only $3\%$ larger. 

\subsection{A Closer Look at Expression $f_4$}
\begin{figure}[htbp]
    \centering
    \includegraphics[width=0.4\textwidth]{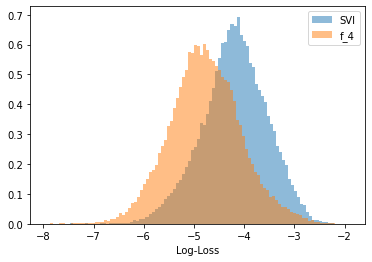}
    \caption{Log-Loss distribution of expression $f_4$ vs. SVI} 
    \label{fig:loss_distribution}
\end{figure}

\begin{figure}[htbp]
    \centering

    \includegraphics[width=\textwidth]{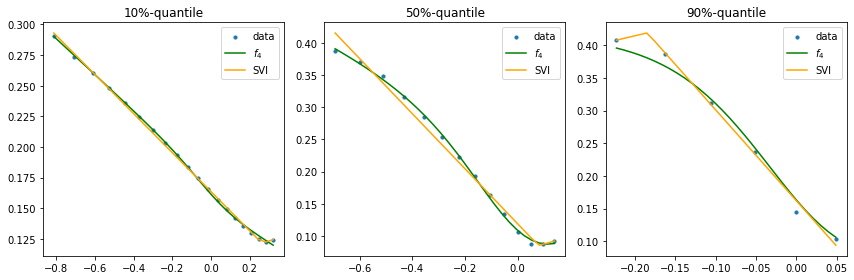}
    \caption{Three fits of the $f_4$-parametrization, representing the $10\%$-, the $50\%$- (median) and $90\%$-quantile in terms of fitting error. The SVI-fit for each slice is also shown.} 
    \label{fig:fits}
\end{figure}

To evaluate the results for the discovered expression
\begin{equation}\label{eq:f4}
f_4(x) = p_4 \left(\tanh(p_1 x) + p_2\right) (p_3 + x) + p_5
\end{equation}
more closely, we have plotted the whole log-loss distribution (i.e. the distribution of log-loss for each separate IV slice) in Figure~\ref{fig:loss_distribution}. We can see that the improvement in terms of log-loss averaged over the sample, as reported in Table~\ref{tab:expr}, is representative of the whole distribution, i.e. the fit to almost every slice is improved with respect to SVI. To get a more refined picture we plot the `quantile fits' for the $q$-quantiles with $q \in \set{10\%, 50\%, 90\%}$ in Figure~\ref{fig:fits}, i.e. the left panel shows the fit whose loss corresponds to the lower decile of all losses, for the middle panel to the median of losses, and for the right panel to the upper decile of all losses. Overall, it seems that with \eqref{eq:f4} symbolic regression has discovered a parametrization that is competitive with, if not superior to, the SVI parametrization of total implied volatility. As a caveat, we note that an analysis of sufficient conditions for the absence of butterfly- and calendar-arbitrage, as performed for SVI in \cite{gatheral2014arbitrage}, is left open for expression $f_4$.

\section{Results on Data Set D2}\label{sec:D2}
The data set D2 is a curated data set consisting of eight IV slices of `SVI5-type', twelve slices of `C6-type' and fourteen slices of `C8-type'. The SVI5-type slices are IV curves that can be fitted almost without error with the 5-parameter SVI model as given in \eqref{eq:SVI}. The C6- and C8-type slices are IV curves with unusual shapes (e.g. $W$-shapes, see also \cite{glasserman2023w}), which can be well-fitted with the proprietary C6- and C8-parametrizations of \texttt{VolaDynamics}, which are unknown to us. Data set D2 is small enough to run conditional symbolic regression on the full data set, without the clustering described in Section~\ref{sec:clustering}. Again, we performed several runs, which led to different expressions due to the random initial conditions of the \texttt{PySR} expression search. 

\subsection{The SVI5 Curves and No-Arbitrage as an Inductive Bias} The SVI5 curves are interesting, because they allow us to evaluate the ability of symbolic regression to re-discover the SVI parametrization from a small sample of IV slices. We have also used them for a controlled ablation experiment on the wing-arbitrage penalization: Table~\ref{tab:SVI5} shows the results from performing a representative conditional symbolic regression run with and without arbitrage penalty. We make several interesting observations: The leaderboard (a) consists mostly of polynomials and variations thereof, which -- due to Weierstrass' theorem -- posses the universal approximation property but violate wing-arbitrage (unless they only contain linear terms). The leaderboard (b) -- for which wing-arbitrage is penalized -- contains SVI-like expressions and indeed rediscovers the SVI parametrization at complexity 15. We observe that even though wing-arbitrage penalization restricts the search space compared to the unpenalized loss, it leads to lower loss at higher complexities.  We conclude that arbitrage penalization serves as a useful inductive bias, nudging the optimization algorithm towards the low-loss regions of the search space.

\subsection{The C6 and C8 Curves} 
\begin{figure}[htbp]
    \centering

    \includegraphics[width = 0.45 \textwidth]{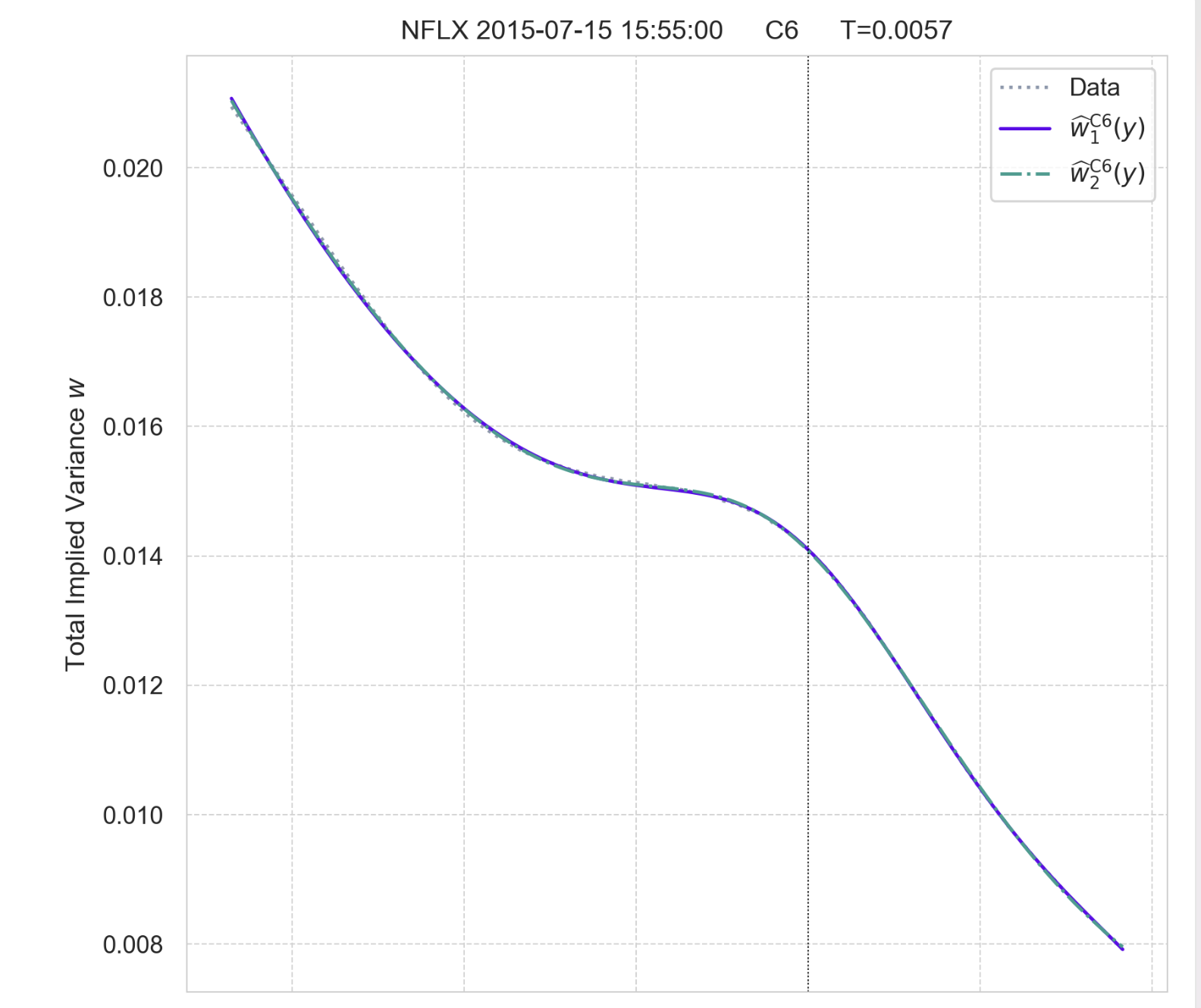}
    \includegraphics[width = 0.45 \textwidth]{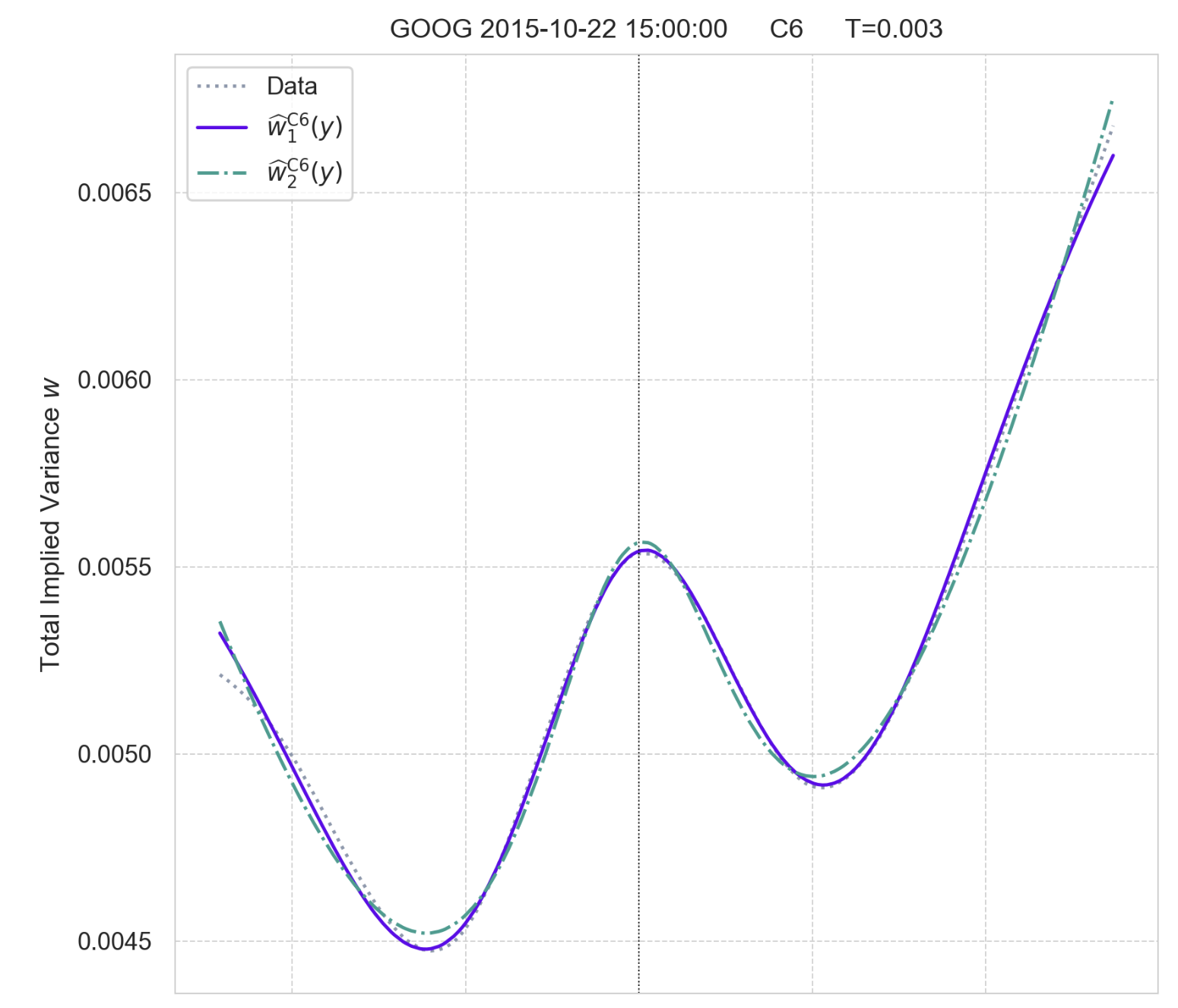}
    \caption{Fit of the discovered parametrizations $\hat w_1^\text{C6}$ and $\hat w_2^\text{C6}$ to two C6-curves from the data set D2. Note that curves and data points are virtually indistinguishable in the left plot.} 
    \label{fig:C6_curves}
\end{figure}

We performed several conditional parametric symbolic regression runs on both the C6-type and the C8-type curves from the data set. For the C6-type curves we found the following two expressions that seem to fit the majority of the available curves well.
\begin{align}
    \label{eq:c6-cand-1}
    \widehat w^\mathrm{C6}_1(x) &= p_1 + p_2 x + p_3 \sqrt{(x - p_4)^2 + p_5^2} + p_6 \log\left(p_7 + (x - p_4)^4\right) \\
    \label{eq:c6-cand-2}
    \widehat w^\mathrm{C6}_2(x) &= p_1 + p_2 x + p_3 \sqrt{(x - p_4)^2 + p_5^2} + p_6 \sqrt{(x - p_7)^2 + p_8^2}
\end{align}
which have $7$ and $8$ parameters, respectively. This is more than the $6$ parameters we would expect given the name of the curve. However, we still obtain reasonably good fits if we set $p_1 = 0$ or $p_2 = 0$. We observe that both parametrizations are extensions to the standard SVI parametrization in the sense that another term is added. Two selected fits are shown in Figure~\ref{fig:C6_curves}. Note that the right plot in the figure represents the worst fit among all C6 curves. 

\begin{figure}[htbp]
    \centering

    \includegraphics[width = 0.45 \textwidth]{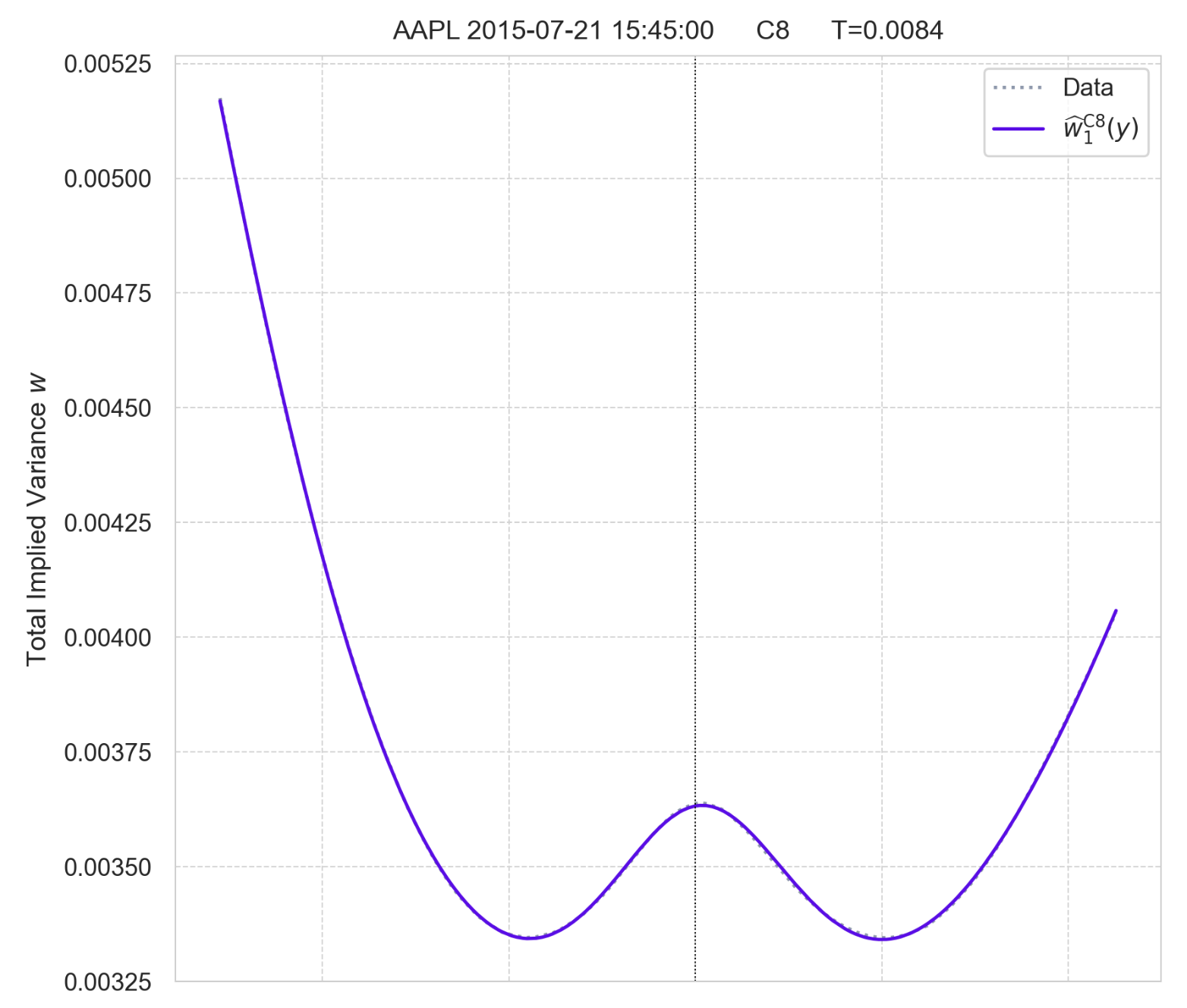}
    \includegraphics[width = 0.45 \textwidth]{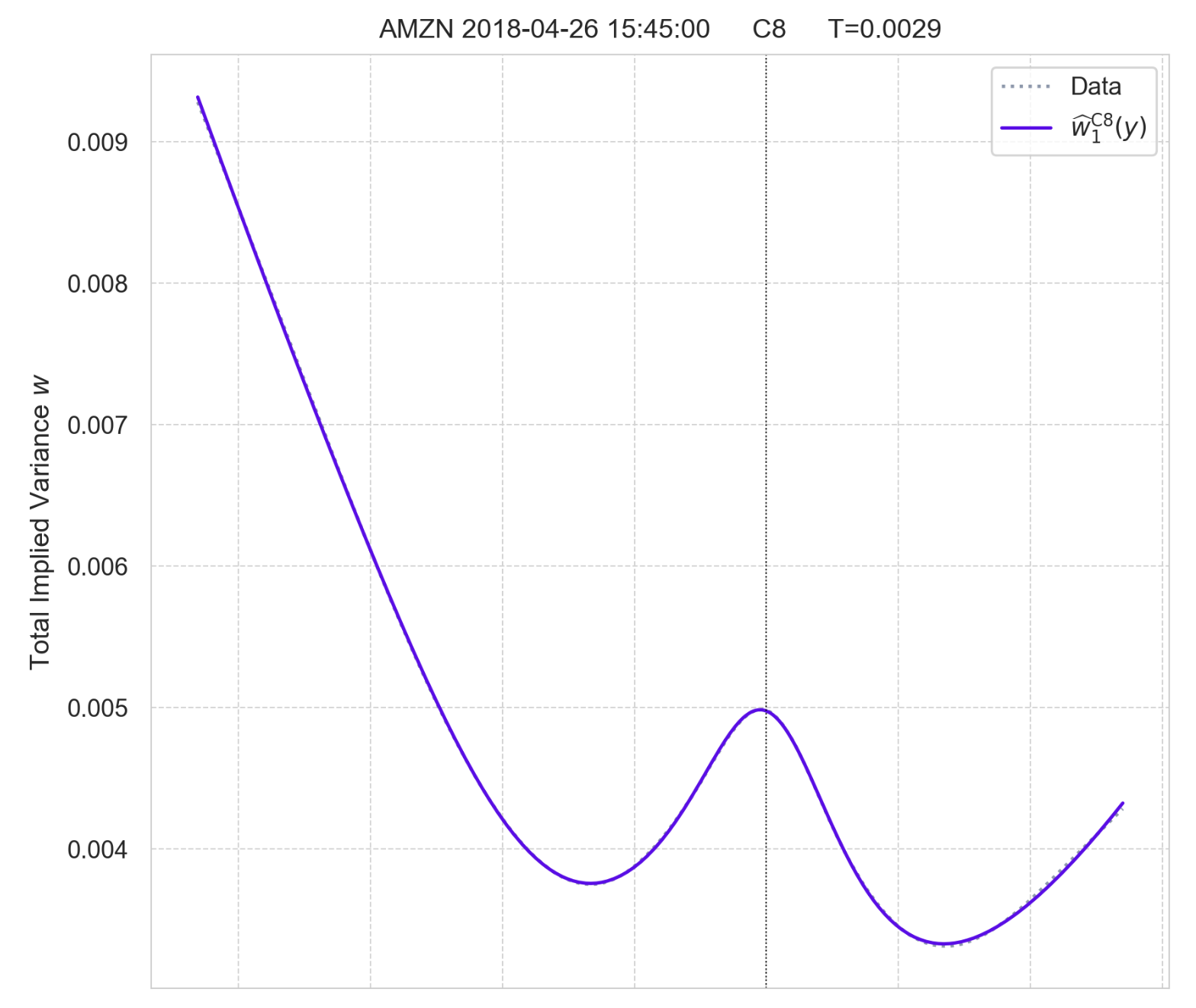}
    \caption{Fit of the discovered parametrizations $\hat w_1^\text{C8}$ to two C8-curves from the data set D2. Note that curves and data points are virtually indistinguishable in both plots.} 
    \label{fig:C8_curves}
\end{figure}

For the C8-type curves the expression discovered by symbolic regression that fitted all curves the best is
\begin{equation}
    \widehat w^\mathrm{C8}_1(x) = p_1 + p_2 x + p_3 \sqrt{(x - p_4)^2 + p_5^2} + \sqrt{x^2 + p_6 + p_7 e^{-(x - p_8)^2}}.
\end{equation}
Once again, the discovered expression is similar to the SVI parametrization with an additional term. Two 
selected plots for this curve are displayed in Figure~\ref{fig:C8_curves}. Note that the fitted candidate function does not visibly deviate from the plot data. 

\begin{table}
\centering
\subcaptionbox{No arbitrage penalty}{
\begin{tabular}{@{}l r r  @{}}
\toprule
Complexity & Loss & Expression\\
\midrule
    $11$  & $1.82 \cdot 10^{-7}$  & $p_{4} + x \left(p_{1} + p_{2} + p_{5} x\right)$ \\
        $12$  & $1.38 \cdot 10^{-8}$  & $p_{4} + \left(p_{1} + p_{5} x\right) \left(p_{2} + x\right)^{2}$ \\
        $13$  & $1.38 \cdot 10^{-8}$  & $p_{4} + x \left(p_{1} + x \left(p_{3} x + p_{5}\right)\right)$ \\
        $14$  & $1.09 \cdot 10^{-8}$  & $p_{4} + \left(p_{1} + p_{3} \sqrt{x^{2}}\right) \left(p_{2} + x\right)^{2}$ \\
        $16$  & $2.01 \cdot 10^{-9}$  & $p_{4} + x \left(p_{1} + \left(p_{2} + x \left(p_{3} x + p_{5}\right)\right)^{2}\right)$ \\
        $17$  & $1.79 \cdot 10^{-9}$  & $p_{4} + x \left(p_{1} + x \left(p_{5} + x \left(p_{2} x + p_{3}\right)\right)\right)$ \\
\bottomrule
\end{tabular}
}\\[1em]
\subcaptionbox{With arbitrage penalty}{
\begin{tabular}{@{}l r r r r @{}}
\toprule
Complexity & Loss & Expression\\
\midrule
    $10$  & $7.55 \cdot 10^{-7}$  & $\sqrt{p_{5}^{2} + \left(p_{3} x + p_{4}\right)^{2}}$ \\
        $11$  & $6.60 \cdot 10^{-9}$  & $p_{2} + p_{4} \sqrt{p_{3} + \left(p_{5} + x\right)^{2}}$ \\
        $13$  & $6.60 \cdot 10^{-9}$  & $p_{2} + p_{4} \sqrt{p_{3} + p_{4} + \left(p_{5} + x\right)^{2}}$ \\
        $\mathbf{15}$  & $\mathbf{4.06 \cdot 10^{-11}}$  & $\mathbf{p_{1} x + p_{2} + p_{4} \sqrt{p_{3} + \left(p_{5} + x\right)^{2}}}$ \\
        $17$  & $4.01 \cdot 10^{-11}$  & $p_{1} x + p_{2} + p_{4} \sqrt{p_{3} + p_{4} + \left(p_{5} + x\right)^{2}}$ \\
\bottomrule
\end{tabular}
}

    \caption{\label{tab:SVI5}Leaderboards (truncated at complexities 10 and 17) for symbolic regression on the SVI5 curve. Table (a) shows results without arbitrage penalty, Table (b) shows results with arbitrage penalty. The rediscovered SVI parametrization is highlighted in bold.}

\end{table}

\section{Discussion and Conclusion}\label{sec:conclusion}
We have demonstrated that symbolic regression is able to discover new parametrizations of implied volatility, which are competitive with SVI in terms of complexity, accuracy and absence of arbitrage. Moreover, symbolic regression is able to rediscover SVI from a small sample of SVI-type slices and discovers simple additive extensions of SVI when run on W-shaped and similar unusual IV slices. From our ablation experiment, it seems that the wing-arbitrage penalty is not only important to avoid asymptotic arbitrage, but that it also serves as an inductive bias that directs the symbolic regression method to low-loss regions of the search space.
For future research it would be interesting to include butterfly-arbitrage as an additional penalty, which is challenging because of the second-order-derivatives that it induces into the loss function. Futhermore, it would be interesting to find symbolic expressions that are able to fit whole surfaces (including the maturity dimension) instead of just slices of implied volatility. 

\footnotesize

\section{Acknowledgement}ChatGPT-5.3 was used for language improvement during preparation of this manuscript.


\bibliographystyle{alpha}
\bibliography{references}

\begin{thebibliography}{HKLW02}

\bibitem[Cra23]{cranmer2023interpretable}
Miles Cranmer.
\newblock Interpretable machine learning for science with {PySR} and
  {SymbolicRegression.jl}.
\newblock {\em arXiv preprint arXiv:2305.01582}, 2023.

\bibitem[Dou12]{doust2012no}
Paul Doust.
\newblock No-arbitrage {SABR}.
\newblock {\em The Journal of Computational Finance}, 15(3):3, 2012.

\bibitem[Fen05]{fengler2005semiparametric}
Matthias~R Fengler.
\newblock {\em Semiparametric modeling of implied volatility}.
\newblock Springer, 2005.

\bibitem[GJ14]{gatheral2014arbitrage}
Jim Gatheral and Antoine Jacquier.
\newblock Arbitrage-free {SVI} volatility surfaces.
\newblock {\em Quantitative Finance}, 14(1):59--71, 2014.

\bibitem[GJMN16]{guo2016generalized}
Gaoyue Guo, Antoine Jacquier, Claude Martini, and Leo Neufcourt.
\newblock Generalized arbitrage-free {SVI} volatility surfaces.
\newblock {\em SIAM Journal on Financial Mathematics}, 7(1):619--641, 2016.

\bibitem[GP23]{glasserman2023w}
Paul Glasserman and Dan Pirjol.
\newblock {W}-shaped implied volatility curves and the gaussian mixture model.
\newblock {\em Quantitative Finance}, 23(4):557--577, 2023.

\bibitem[HKLW02]{hagan2002managing}
Patrick~S Hagan, Deep Kumar, Andrew~S Lesniewski, and Diana~E Woodward.
\newblock Managing smile risk.
\newblock {\em The Best of Wilmott}, 1:249--296, 2002.

\bibitem[Lee04]{lee2004moment}
Roger~W Lee.
\newblock The moment formula for implied volatility at extreme strikes.
\newblock {\em Mathematical Finance: An International Journal of Mathematics,
  Statistics and Financial Economics}, 14(3):469--480, 2004.

\bibitem[MM22]{martini2022no}
Claude Martini and Arianna Mingone.
\newblock No arbitrage {SVI}.
\newblock {\em SIAM Journal on Financial Mathematics}, 13(1):227--261, 2022.

\bibitem[Rop09]{roper2009implied}
Michael Paul~Veran Roper.
\newblock {\em Implied volatility: General properties and asymptotics}.
\newblock PhD thesis, UNSW Sydney, 2009.

\bibitem[VC24]{vuletic2024volgan}
Milena Vuleti{\'c} and Rama Cont.
\newblock Volgan: a generative model for arbitrage-free implied volatility
  surfaces.
\newblock {\em Applied Mathematical Finance}, 31(4):203--238, 2024.

\bibitem[{Vol}25]{voladynamics}
{Vola Dynamics LLC}.
\newblock Webpage voladynamics.com.
\newblock Accessed 2025-05-26, 2025.

\end{thebibliography}

\end{document}